\documentstyle[11pt,aaspp4]{article}

\begin{document}

\lefthead{Makino and Ebisuzaki}
\righthead{Merging of Galaxies}

\title{Merging of galaxies with central black holes \\
II. Evolution of the black hole binary and the structure of the core}

\author{Junichiro Makino}

\affil{Department of Graphics and Computer Science,\\
College of Arts and Sciences, University of Tokyo,\\
3-8-1 Komaba, Meguro-ku, Tokyo 153, Japan.}

\def\undertext#1{{$\underline{\hbox{#1}}$}}
\def\sub#1{_{{\rm #1}}}
\def\doubleundertext#1{{$\underline{\underline{\hbox{#1}}}$}}
\def\half{{\scriptstyle {1 \over 2}}}
\def\ie{{\it {\frenchspacing i.{\thinspace}e. }}}
\def\eg{{\frenchspacing e.{\thinspace}g. }}
\def\cf{{\frenchspacing\it cf. }}
\def\etal{{\frenchspacing\it et al.}}
\def\et{{\etal}}
\def\simlt{\hbox{ \rlap{\raise 0.425ex\hbox{$<$}}\lower 0.65ex\hbox{$\sim$} }}
\def\simgt{\hbox{ \rlap{\raise 0.425ex\hbox{$>$}}\lower 0.65ex\hbox{$\sim$} }}
\def\solar{\odot}
\def\msun{\ifmmode{M_\solar}\else{$M_\solar$}\fi}
\def\rsun{\ifmmode{R_\solar}\else{$R_\solar$}\fi}
\def\Rf{\parindent=0pt\smallskip\hangindent=3pc\hangafter=1}
\def\pc{{\rm pc}}
\def\kpc{{\rm kpc}}
\def\Mpc{{\rm Mpc}}
\def\yr{{\rm yr}}
\def\Myr{{\rm Myr}}
\def\Gyr{{\rm Gyr}}
\def\kT{\ifmmode{kT}\else{$kT$}\fi}
\def\N{{\ifmmode{N}\else{$N$}\fi}}
\def\fb{{\ifmmode{f_B}\else{$f_B$}\fi}}
\def\emax{{\ifmmode{E_{max}}\else{$E_{max}$}\fi}}
\def\td{{\ifmmode{t_d}\else{$t_d$}\fi}}
\def\tcr{{\ifmmode{t_{cr}}\else{$t_{cr}$}\fi}}
\def\tr{{\ifmmode{t_r}\else{$t_r$}\fi}}
\def\trh{\ifmmode{t_{rh}}\else{$t_{rh}$}\fi}
\def\vv{{\ifmmode{\langle v^2\rangle}\else{$\langle v^2 \rangle$}\fi}}
\def\v{{\ifmmode{\langle v^2\rangle^{1/2}}
		\else{$\langle v^2 \rangle^{1/2}$}\fi}}
\def\half{{\ifmmode{{1 \over 2}}\else{${1 \over 2}$}\fi}}
\def\dhalf{{\textstyle {1 \over 2}}}
\def\threehalf{{\ifmmode{{3 \over 2}}\else{${3 \over 2}$}\fi}}
\def\dthreehalf{{\textstyle {3 \over 2}}}
\def\dfivehalf{{\textstyle {5 \over 2}}}
\def\dfivethree{{\textstyle {5 \over 3}}}
\def\kms{\ifmmode{\rm km\,s^{-1}}\else{$\rm km\,s^{-1}$}\fi}
\def\kmps{{\rm km/s}}

\begin{abstract}

We investigated the evolution of the black hole binary formed by the
merging of two galaxies each containing a central massive black
hole. Our main goal here is to determine if the black hole binary can
merge through the hardening by dynamical friction and the
gravitational wave radiation. We performed $N$-body simulations of
merging of two galaxies with wide range of total number of particles
to investigate the effect of the number of particles on the evolution
of the black hole binary. We found that the evolution timescale was
independent of the number of particles in the galaxy $N$ until the
separation reaches a critical value. After the separation became
smaller than this critical value, the evolution timescale was longer
for larger number of particles. Qualitatively, this behavior is
understood naturally the result the ``loss-cone'' effect. However, the
dependence of the timescale on $N$ is noticeably weaker than the
theoretical prediction. In addition, the critical separation is
smaller than the theoretical prediction. The timescale of evolution
through gravitational radiation at this critical separation is longer
than the Hubble time.  We discuss the reason of these discrepancy and
the implication of the present result on the structure of the
ellipticals and QSO activities.

\end{abstract}

subject headings: galaxies:interactions, galaxies:nuclei.

\newpage

\section{Introduction}\label{sec:intro}

The possibility of the formation of massive black hole binaries at the
cores of the elliptical galaxies was first pointed out by Begelman, et
al. (1980, hereafter BBR). Their argument is summarized as follows. The
most likely energy source for AGN or QSO activities is the massive
black holes with the mass $M\sub{BH} \simeq 10^8 M\sub{\odot}$. If two
galaxies with central massive black holes merge with each other,
the BHs will sink to the center of the merger remnant because of the dynamical
friction from stars, and form a binary. 

BBR argued that such a binary would have the typical lifetime larger
than the Hubble time,  since the BH binary would create the
``loss cone'' in the distribution of field particles, which is
repopulated only in the relaxation timescale of the core.
According to their argument, once the separation of the binary comes
down to the loss-cone radius, $r\sub{lc}$, which is given by
\begin{equation}\label{eq-defrlc}
   r\sub{lc} = \left(M\sub{BH} \over M\sub{c}\right)^{3/4} r\sub{c},
\end{equation}
the evolution timescale of the binary becomes the thermal timescale of
the core. Here, $M\sub{BH}$ is the mass of a black hole, $M\sub{c}$ is
the mass of the core, and $r\sub{c}$ is the core radius.
When the separation between the black holes becomes smaller than this
$r\sub{lc}$, the binary effectively sweeps out the stars that can interact
with the binary. Thus the evolution timescale becomes the thermal
timescale of the core. They concluded that the black hole binary in
the merger remnant has a lifetime much longer than the Hubble time.

Ebisuzaki et al. (1991) pointed out that the lifetime is much shorter
if the BH binary is highly eccentric.  BBR assumed that the binary
remains circular. The timescale of the gravitational radiation is
proportional to $(1-e)^{-3.5}$ for $e\sim 1$, where $e$ is the
eccentricity. If the BH binary becomes highly eccentric, therefore,
the lifetime would become much shorter. In fact, even for the average
``thermal'' eccentricity of 0.7, the lifetime is shorter than that of
a circular binary by nearly two orders of magnitudes.  The binary
evolves through the dynamical friction from field particles.  If we
naively apply Chandrasekhar's dynamical friction formula, the
eccentricity should grow quickly since the dynamical friction is
inversely proportional to the third power of the velocity, and
therefore the strongest at the apoastron (\cite{Fukushige1992}).

Mikkola and Valtonen (1992) numerically integrated the evolution of a
BH binary in the core of a galaxy. They followed the orbit of BHs in
the distribution of the field stars. The gravitational interaction
between BHs and those between field stars and BHs are directly
calculated. The interaction between the field stars is expressed by
the force from a fixed potential. As a result, they could not follow
the evolution of the structure of the core. However, they could use a
fairly large number of particles ($N=10,000$). They found that the
binary hardened linearly in time. In other words, in their calculation
the loss-cone depletion did not take place. For the eccentricity, what
they found is that the eccentricity increases only very slowly. The
rapid increase of the eccentricity predicted by Fukushige \etal\ (1992)
did not take place. This is not very surprising because the standard
dynamical friction formula used by Fukushige \etal\ (1992) is valid only for
the motion of a single massive particle, and therefore is not guaranteed
to give a correct result for the evolution of a binary.  They also
determined the rate of evolution of energy and eccentricity of the BH
binary through scattering experiment. The result of their scattering
experiment was consistent with the result of simulation.

Quinlan (1996) performed extensive scattering experiments of a BH
binary and a field particle. He extended the work of Mikkola and
Valtonen (1992) to include unequal-mass black holes. His result is
similar to that of Mikkola and Valtonen (1992). The change in the
eccentricity is small unless it is initially close to unity.

Makino \etal\ (1993) made self-consistent $N$-body simulation of the
evolution of a BH binary in the core of a galaxy with 16,384
particles. Their results are summarized as follows: (a) The BH binary
becomes harder beyond the loss-cone radius, without showing any sign
of slowing down at least to $0.1r_{\rm lc}$, where simulations were
stopped.  (b) After the binary is formed, the eccentricity of the
binary remains roughly constant.  (c) The eccentricity depends
strongly on the initial condition of the BHs. Thus, their result is
again consistent with the result of Mikkola and Valtonen (1992), but
apparently in contradiction with the theoretical prediction of BBR.

In the present paper, we give the result of self-consistent direct
$N$-body simulations with the number of particles much larger than
that was employed in Makino \etal\ (1993) or Mikkola and Valtonen
(1992).

In order to study the evolution of the binding energy of the central
BH binary, we performed the merging simulations from the same initial
condition, but with several different number of particles. The
small-$N$ effect decreases the evolution timescale of the BH binary
by at least two different ways.  The first is that the
timescale of the repopulation of the loss cone is proportional to the
two-body relaxation time of the core. Therefore, if the timescale of
the evolution of the binding energy is actually determined by the
repopulation timescale, it should be proportional to the core
relaxation time.  In $N$-body simulations, however, the core
relaxation time is not much larger than the timescale of the depletion
itself. If the repopulation is faster than the depletion, the growth
rate of the BH binary would be  independent of $N$.

The other reason is that the center of mass of the BH binary has a
small random velocity, since its kinetic energy is in equipartition
with those of field particles (Bahcall and Wolf 1976, Mikkola and
Valtonen 1992).  Thus the BH binary might wander in, or even to
outside, the loss cone. As a result, the effective radius of the loss
cone might become larger in $N$-body experiments. The random velocity
would be proportional to the square root of the mass ratio between the
field particles and a BH binary. Therefore, a BH binary in the
numerical simulation has the random velocity several hundred times
larger than that of a real BH binary.  Note that this second problem
exists even in simulations with fixed field potential such as
performed by Mikkola and Valtonen (1992), since in these calculations
BHs feel the forces from the field particles, which shows the same
fluctuating behavior as in self-consistent $N$-body
calculations. Moreover, the effect of the relaxation is not completely
suppressed either, because the orbits of field particles are changed
by the interaction with the BH particles as well. Though the field
particles do not interact directly with one another, they can still
indirectly interact through interaction with BH particles.

In section 2, we describe the numerical method we used.  In
section 3, we describe the result of the simulations with different
numbers of particles.  Our main result is that the evolution timescale
of the binary depends on the total number of field particles, $i.e.$,
the ratio between the mass of BH, $M_{BH}$ and the mass of field
particles, $m_{field}$. The relation between the timescale of the
evolution of the binding energy of the BH binary, $t_b$, and the mass
ratio between the BH and field particles is expressed as
\begin{equation}
t_{b} \propto \left({M_{BH}\over m_{field}}\right)^{0.3},
\end{equation}
after the separation of two BHs reached a critical value which depends
on the initial structure of the core.  While the separation is larger
than this critical value, the evolution timescale is almost independent of
$N$. This critical value can be qualitatively understood as the loss
cone radius, but is much smaller than the prediction of equation
(1). The more sophisticated treatment by Quinlan (1996) seems to give
better understanding.

For the timescale, a naive theoretical estimate based on the
relaxation timescale should give $t_{b} \propto ({M_{BH}/m_{field}})$
after the separation becomes smaller than the critical (loss-cone)
separation.  Section 4 is for discussions. 

\section{NUMERICAL METHOD AND INITIAL CONDITION}

\subsection{NUMERICAL METHOD}

The calculation code we used is NBODY1 (\cite{Aarseth1985}) modified
to be used with GRAPE-4 special purpose computer for gravitational
$N$-body simulation (\cite{Taiji1996}).  The time integration scheme
is changed to the Hermite scheme (\cite{Makino1992}) to take advantage
of the GRAPE-4 hardware. We neglect relativistic effects and treated
BH particles as massive Newtonian particles. This treatment is good as
long as the periastron distance of the BH pair does not become very
small.  Our simulations were terminated well before the relativistic
effect would become important.  We used a softened potential with the
softening of $1/1024$ for the force between field particles, and pure
$1/r$ potential for forces to and from BH particles.  The relative
accuracy of the force calculated on GRAPE-4 is about 7 digits (IEEE
single precision), which is more than enough for forces from field
particles but might not be sufficient for the force from BH particles.
In the present code, the force from field stars are calculated on
GRAPE-4, while that from BH particles are calculated on the host
computer in full double precision.

Calculations were performed on one cluster of GRAPE-4 system with the
theoretical peak speed of 270 Gflops. Actual sustained speed was around
100-150 Gflops for simulations with 128K-256K particles. The 256k run
took about five CPU days. 

\subsection{Initial Conditions}

The procedure to prepare the initial condition is the same as
described in Makino and Ebisuzaki (1996, hereafter referred to as
Paper I). We used the King model with the nondimensional central
potential $W_c$ of 7 as the initial galaxy model.  The number of
particles is 2048 to 262144.  The initial galaxy is created so that
the total mass $M_g$ is one and the total energy $E_g$ is $-1/4$ in
the system of units where gravitational constant $G$ is one (the
standard unit, \cite{Heggie1986}). The mass of a field particle is
$m_{field} = 1/N$.  The half mass radius of the initial galaxy model
is about 0.75 and the half mass crossing time is $2\sqrt{2}$. To place
the central BH, we removed $M_{BH}/m_{field}$ particles closest to the
center of the galaxy at time $t=0$ and put the BH particle at the
center. Thus the galaxy do not initially have a strong central cusp.
In Paper I, we found that the structure of the merger does not depend
on the details of the procedure to place the BH particles. For all
calculations, $M_{BH}=1/32$ unless otherwise specified.

The initial orbit of two galaxies is parabolic with the
periastron distance equal to 1. The initial separation of two galaxies 
is 10. We integrated the system to time $t=60$.  

\section{Results}

\subsection{The evolution of the BH binary.}

Figure 1 shows the time evolution of the energy of the BH binary,
$E_b$, for all runs. The energy $E_b$ is defined as
\begin{equation}
E_b= {1 \over 2}\mu v_b^2 - {M_{BH}^2 \over r_b} = - {M_{BH}^2 \over
2a},
\end{equation}
where $\mu = M_{BH}/2$ is the reduced mass of the BH binary, $r_b$ and
$v_b$ are the relative distance and velocity of the two BHs, and $a$
is the semi-major axis of the BH binary.

Before two galaxies merge, BHs lie at the centers of galaxies. Thus
the distance is large and the binding energy of two BHs is negligible. 
In this period, $E_b$ is roughly equal to the sum of the kinetic
energies of two BHs. The energy $E_b$ takes a maximum around $t=13$
because at that time two galaxies pass the periastron. The relative
velocity of two galaxies is the largest at the periastron. Therefore
$E_b$ takes the maximum value.  By $t \simeq 30$, the second fallback
takes place and  two galaxies merge. Soon after that, two BHs form
a binary.

\placefigure{fig1}

From figure 1 we can see that the hardening rate of binaries,
$-dE_b/dt$, is quite different for runs with different number of
particles. The hardening rate is smaller for larger $N$. If we
investigate figure 1 more closely, it seems the difference in the
hardening rate is larger for the later time. Until $E_b$ reaches
around $-0.05$, the growth rate seem to be similar for all runs except
for $N=2048$. In the 2048-body run, one of the BH particles had formed
a bound pair with a field particle before it has become bound with
another BH. When two BHs became bound, this field particle was still
bound to a BH particle. This particle was bound to the BH binary until
$t=40$. This is the reason why the curve for the 2048-body run is very
noisy.

In order to see the dependence of the hardening rate on the energy
itself, we plotted the hardening rate $-dE_b/dt$ as a function of $N$
in figure 2. Here, $-dE_b/dt$ is calculated for two intervals of
$E_b$, $(-1/160,-1/80)$ and $(-1/10,-1/5)$. Note that for some runs
$E_b$ at the end of the run shown in figure 1 did not reach
$-1/5$. For these runs, we extended simulation so that $|E_b| > 0.2$
at the end of the run.

Here we can clearly see that the hardening rate at the early phase the
evolution seems to converge to $-dE_b/dt \sim 0.008$ for large $N$,
while that for later phase of the evolution shows the power-law like
behavior expressed roughly as $dE_b/dt \propto N^{-1/3}$ for the
entire range of $N$.

\placefigure{fig2}

Figure 3 shows the evolution of $E_b$ for repeated 256k runs. Here,
the final merger of one run is used as the initial galaxy of the next
run, following the procedure described in Paper I. In these runs, the
central density of the initial galaxy becomes somewhat higher as we
repeat the merging process. (See figure 4d for the density
profiles). Since the timescale of the evolution of the binary is
proportional to the central density, this difference in the central
density resulted in the difference in the growth timescale. This
difference in the central density also resulted in the difference in
the critical value of $E_b$.

\placefigure{fig3}

There is no reason to assume that the first merger would give a
reasonable result for the timescale, since our choice of the initial
model is rather arbitrary.  The structure of the core in the last
merger events would be a more appropriate model for the initial
galaxy.  In this case, the critical value for $E_b$ seems to be around
$-0.3$.

\subsection{The structure of the core}

Figure 4 shows the density profiles of the central region of the
merger remnant from 256k, 32k and 4k runs for different times. For
the 256K run, the density is almost flat at the very central region,
with possible minimum at $r\sim 0.01$, where $r$ is the radius. For
32k runs, the central region has a cusp of roughly $\rho \sim
r^{-0.5\sim -1}$, and for 4k $\rho \sim r^{-1\sim -1.5}$. For all
runs, central density decreases in time.

\placefigure{fig4}

The difference in the central structure explains the difference in the
growth rate of $E_b$ shown in figures 1 and 2. The growth rate is
proportional to  the stellar mass density around the BH
binary. For higher $N$ the density is lower. Therefore growth rate is
smaller for larger $N$. 

The difference in the structure of the core is due to the difference
in the central relaxation time. Since the merger remnant does not have 
a flat core, the central relaxation time is rather difficult to
define. We calculated the local relaxation time as the function of the 
radius $r$, and found that it takes the minimum value around $r=0.1$
for all models. So we regard this minimum value as the central
relaxation time, $t_{rc}$. At time $t=40$, $t_{rc}$ is
around 200, 800 and 5000 for 4k, 32k and 256k runs,
respectively. Thus, for the 4k run the simulation timespan is comparable
to the relaxation time, while for the 256k run the relaxation timescale is 
orders of magnitudes larger than the simulation timespan. 

The flat central region observed in the 256k run is a kind of the loss
cone predicted by BBR. In BBR's theory, when the binary becomes
sufficiently hard, it kicks out all the stars it can interact and then
the growth slows down significantly.

It, however, should be noted that the structure we observed is rather
different from BBR's picture or the standard model for the star
cluster with massive central BH (\cite{Shapiro1985}, and references
therein), which predict the cusp with $\rho \propto r^{-1.75}$ outside
a few times the radius of the BH binary. For the 256k run, the
numerical result is a very shallow cusp outside the loss cone like
$\rho \propto r^{-0.5}$. For runs with smaller $N$ the slope is
steeper. Thus, it seems the cusp becomes shallower as we increase $N$. 
On the other hand, the theoretical prediction is a universal cusp of
$\rho \propto r^{-1.75}$.  We will discuss possible interpretations in
section 4.

Figure 5 gives the surface brightness profiles for same runs as in
figure 4. Here, the time evolution is not as pronounced as in figure
3. In addition, we cannot see any decrease of the luminosity toward
the center even for the 256k run.  

\placefigure{fig5}

\section{Summary and Discussion}

We performed the simulation of the evolution of massive BH binaries
formed through mergings of ellipticals by means of direct $N$-body
simulations. Our major findings are summarized as follows.

\begin{description}

\item{a)} The timescale of the evolution of the binary depends on the
          total number of particles, in other words, the mass ratio
          between BHs and field particles. However, the dependence is
          very weak for the early phase of evolution. For later phase,
          we observed $t_b \propto N^{1/3}$, which is much weaker than
          the prediction of the theoretical model.

\item{b)} The evolution timescale shows very strong dependence on the
          initial central density. 

\item{c)} For large $N$ runs, the ``loss cone'' depletion effect is
          clearly visible. In other words, the stellar density around
          the BH is lower for larger $N$.  However, again, the density
          profile we obtained for simulations with large $N$ was quite
          different from the theoretical model which gives the density
          cusp of $\rho \propto r^{-1.75}$.

\end{description}

In the following, we discuss the implication of these results in some
details.

\subsection{The hardening rate of the BH binary}

For the early phase of the evolution, it is not surprising that the
evolution timescale is independent of $N$. This is because the loss
cone depletion has not taken place.  However, our result shows that
the the loss-cone radius is much smaller than the prediction by BBR.

This difference might be because of BBR's assumption in calculating the
loss-cone radius. They assumed that the loss-cone depletion occurs
when the total kinetic energy of all stars which can interact with the
binary becomes smaller than the binding energy of the binary. Quinlan
(1996) argued that the loss-cone depletion occurs only when all
particles that can interact with the BH binary are ejected out of the
core. In his examples, this assumption leads to the loss-cone radius a
few times smaller than the semi-major axis for the ``$1kT$'' binary
(the binary with the orbital velocity comparable to the typical
velocity of field stars). His result is qualitatively consistent with
our present result. For our runs, the critical value of $E_b$ is $-2\sim 10
 \cdot M_{BH}v_c^2$, depending on the density profile, where
$v_c$ is the r.m.s. velocity of field particles. 

After the BH reaches the loss cone radius, its evolution slowed
down. However, the dependence of the growth timescale on $N$ is much
weaker than the theoretical prediction of $t_b \propto N$. This
difference might be understood by means of the detailed treatment of
the evolution of the stellar distribution around the central black
hole developed by Shapiro and his collaborators (\cite{Shapiro1985},
\cite{Duncan1983}).

The essence of their argument is that the stars
are removed from the system only when it actually come close enough to
the BH (they considered the distribution of stars around a single BH). 
Thus, if the average change of the angular momentum of a typical star
in the core in one orbital period is larger than the maximum angular
momentum of the star in the loss cone orbit, the loss cone is
repopulated in the timescale shorter than the depletion timescale. In
this case, the dependence of the growth timescale on $N$ might be
weaker than the theoretical prediction.

The maximum angular momentum of a field particle, $J_{max}$, that can
interact with the BH binary is expressed roughly as
\begin{equation}
J_{max} \sim \sqrt{M_{BH} a} \sim 0.01 E_b^{-1/2},
\end{equation}
where $a$ is the semi-major axis of the BH binary an we used
$M_{BH}=1/32$. 

On the other hand, the average change in the angular momentum of
a typical star in the core per orbit is given roughly by 
\begin{equation}
\Delta J \sim r_c v_c \sqrt{t_c/t_{rc}},
\end{equation}
where $r_c$, $v_c$, $t_c$ and $t_{rc}$ are the core radius, the r.m.s. 
velocity of the particles in the core, the crossing time and the
relaxation time of the core, respectively. If we use $r_c = 0.1$ and
$N_c = N/20$ as a typical value for our simulation (see figure 4), we
have
\begin{equation}
\Delta J \sim \sqrt{2 \log(0.02N) \over N}.
\end{equation}

For $E_b \sim -0.1$, $\Delta J > J_{max}$ in the case of  $N=2048$ and the
opposite in the case of $N=262144$. 
Thus, the range of $N$ in our simulation just cover the transition from
$J_{max} < \Delta J $ to $J_{max} > \Delta J $. It is not at all
surprising that the dependence of the growth rate is weaker than the
prediction of the theory which assumes $J_{max} >> \Delta J $. In
order to really determine the dependence of the lifetime on $N$, we
need to employ more particles.

Simulations with number of particles larger than used here might be
possible by means of more approximate schemes such as the treecode
with individual timesteps (McMillan and Aarseth 1993) or an extension of
the SCF method (\cite{Quinlan1995}) to
the case of central BH binary.

It should be noted that here we study the {\it thermal} evolution of
the system since if the system is collisionless the loss cone will
never be repopulated. Thus, the two-body relaxation effect should be
modeled appropriately.  The use of the so-called ``collisionless''
scheme should be done very carefully.  

\subsection{The evolution of BH binaries in real ellipticals}

Here we discuss the evolution of BH binaries in real ellipticals.  For
simplicity, we assume that the evolution of the binary stops at the
critical energy obtained by our experiments.

The timescale of the
gravitational radiation is given by
\begin{equation}\label{eqn:tgr0}
t\sub{GR} =2\times 10^{15} g(e)  (M\sub{BH}/10^8M\sub{\odot})^{-3}
            (a/1 {\rm pc})^{4} \quad{\rm yr},
\end{equation}
with $g(e)$ given by
\begin{eqnarray}\label{eqn:ge}
g(e) &=& {(1-e^2)^{7/2} \over 1+(73/24)e^2 + (37/96)e^4}\nonumber\\
     &\simeq& 2(1-e)^{7/2} \quad (e \sim 1).
\end{eqnarray}
which, in our unit,
%
\begin{equation}
t\sub{GR} =5\times 10^{9} g(e)  (M\sub{BH}/10^8M\sub{\odot})
            (v_c/300 {\rm km/s})^{-8} E_b^{-4}\quad{\rm yr}.
\end{equation}
Thus, the lifetime of the BH binary with $E_b = -0.05$ is
${\sim 10^{15}}$ years,  which is  well over the Hubble time. For 
$E_b=-0.3$, the lifetime is more than $10^{11}$ years. 
Unless the eccentricity is relatively large, it is unlikely 
for the BH binary to merge. 

BBR (see also Rees 1990) argued that if the BH binary does not merge
in the time significantly shorter than the Hubble time it is likely to
be kicked out from the core or the galaxy itself when the parent
galaxy merges with yet another galaxy, or at least kicks out the third
binary, through the gravitational slingshot. 

However, the slingshot mechanism is not as effective as assumed by
BBR. When one BH binary and a single BH reside in the core, there are
two possible outcomes. In one case, one of the three BHs is ejected
out of the parent galaxy through the slingshot. In the other case, the
binary merges through gravitational radiation before kicking out the
third BH.  BBR concluded that the slingshot is the likely outcome,
assuming that the binary is circular. Makino and Ebisuzaki (1995)
showed that the slingshot is unlikely to occur if the distribution of
the eccentricity is correctly taken into account. They showed that the
typical lifetime of the BH binary before the ejection of the third
body is less than $10^9$ years. If the BH binary merged before the
third BH is ejected out of the galaxy, the final state is the binary
of the merged BH and the third BH.

Thus, it is quite possible that a large fraction of ellipticals have
central BH binaries. The implication of the presence of the binary was
addressed briefly in BBR. Of course, if the velocity dispersion is
higher, the timescale of the gravitational radiation at the loss cone
separation become much shorter. The opposite is also true. If $v_c <
200{\rm km/s}$, the mechanism described in Makino and Ebisuzaki (1995)
cannot make the BH binary merge before it eject the third body. On the
other hand, if $v_c > 600{\rm km/s}$, the BH binary would merge
through the dynamical friction and the gravitational radiation. Since
most of ellipticals fall in the range of $ 200 {\rm km/s} < v_c <
600{\rm km/s}$, the fate of the BH binary is difficult to
predict. Interaction with gas might determine the final fate.

If triple BH systems are not very uncommon, we might be able to
observe some of them as the QSO activities located far from the center
of the galaxy. Recent HST observations of nearby QSOs (Bahcall,
Kirhakos, and Schneider 1995a, 1995b, 1995c) suggested many of the
QSOs are not located at the center of the galaxy. In the triple BH
system, BHs would spend most of the time in the halo ``parking''
orbit, as in the case of the binaries in globular clusters (\cite{Hut1992}). Thus, QSO activities might be observed in the positions far from
the center of the parent galaxy. To explain QSOs outside the core of
the parent galaxy, Fukugita ant Turner (1996) proposed a scenario in
which the massive BH are formed independently from the galaxies. Such
an extraordinary model might not be necessary if triple BHs are common.

\subsection{The structure of the central region}

Our numerical result for the density profile around the BH pair is
quite different from the standard picture of the cusp with $\rho
\propto r^{-7/4}$. Though the cusp exists, its slope is smaller for
large $N$.  

For large-$N$ runs, it is quite natural that the slope of the cusp is
different from the theoretical prediction, simply because the duration
of the simulation is much shorter than the thermal relaxation time of
the core. It takes the thermal relaxation time of the core for the
cusp to fully develop.  For the 256k run, the local two-body
relaxation time at $r=0.1$ is $5\times 10^3$ for $t=40$. Thus the cusp
cannot develop fast enough.

Since the two-body relaxation has even smaller effect in real
ellipticals than in our largest simulations, we can safely predict
that the $r^{-7/4}$ cusp does not present in ellipticals with central
massive BH. The structure is most likely to be a cusp of $r^{-0.5}$
formed by the particles not bound to the BH binary
(\cite{Duncan1983}), terminated at the loss cone radius. This is in
good agreement with the resent HST observations of the core of large
ellipticals which shows the cusp with the slope $-0.5 \sim -1$.  This
shallow cusp indicates that the central BH is placed in the galaxy in
dynamical timescale. If the BH mass grows in the timescale longer than the
dynamical time, the slope would be $-1.5$ (Young 1980).

\acknowledgements

I'm grateful to Gerry Quinlan for his critical comments on the very
early version of this paper and stimulating discussions.  I also thank
Piet Hut for helpful discussions,  Yoko Funato and Toshiyuki
Fukushige for comments on the manuscript.  This work was supported by
Grant-in-Aid for Specially Promoted Research (04102002) of the
Ministry of Education, Science and Culture, Japan.

{}

\newpage
\section*{Figure Captions}

\figcaption[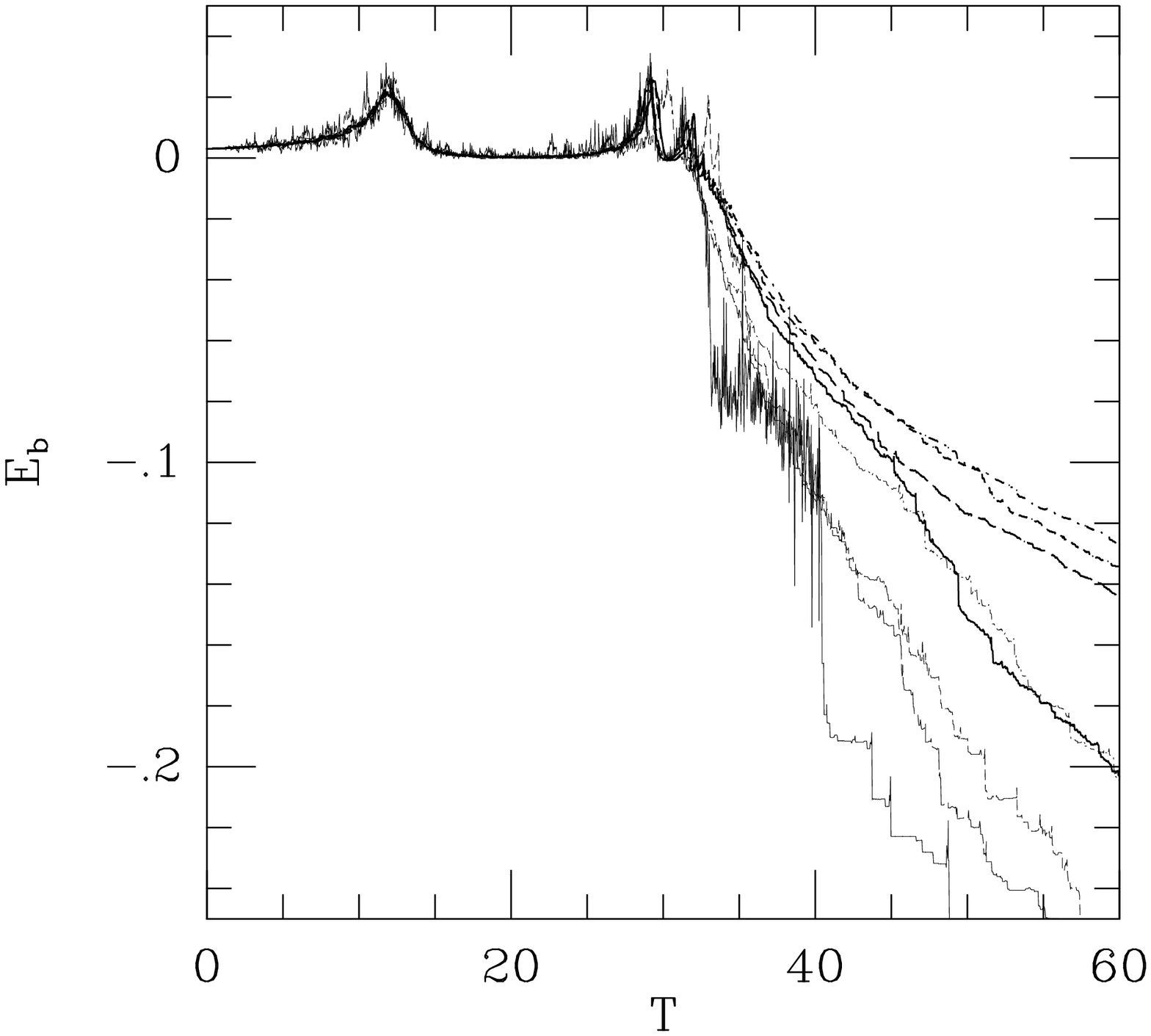]{The time evolution of the energy of the BH
binary for runs with $N=2{\rm k}$ (bottom) to 256k (top). Thin curves
represent 2k (solid), 4k (short dashed), 8k (long dashed) and 16k (dot 
dashed) runs.  Thick curves
represent 32k (solid), 64k (short dashed), 128k (long dashed) and 256k (dot 
dashed) runs. 
\label{fig1}}

\figcaption[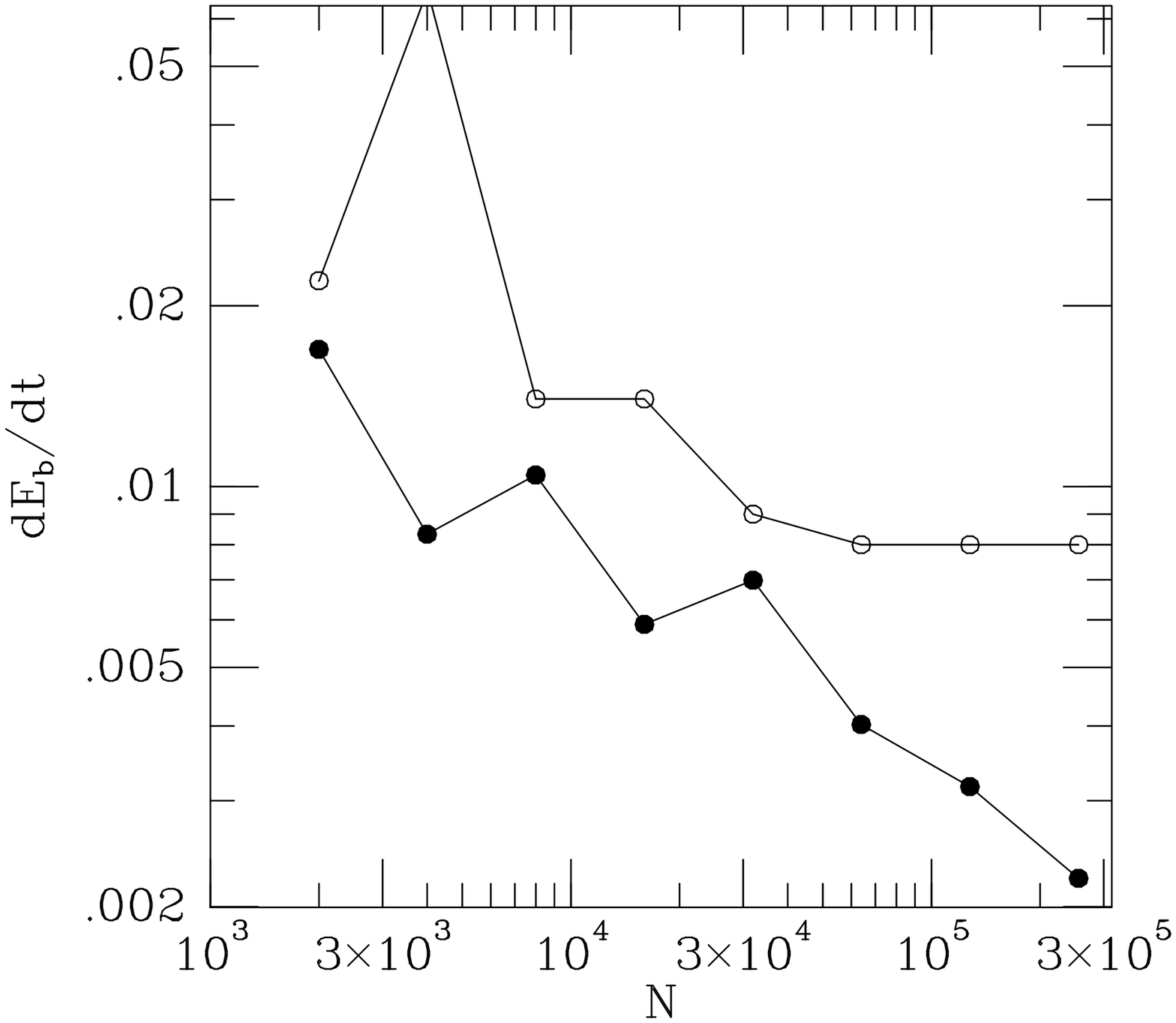]{
The growth rate of the binary binding energy $-dE_b/dt$ plotted versus 
the total number of particles $N$. Open and filled circles the average
rate for $-1/160 > E_b > -1/80$ and $-1/10 > E_b > -1/5,$
respectively. 
\label{fig2}}

\figcaption[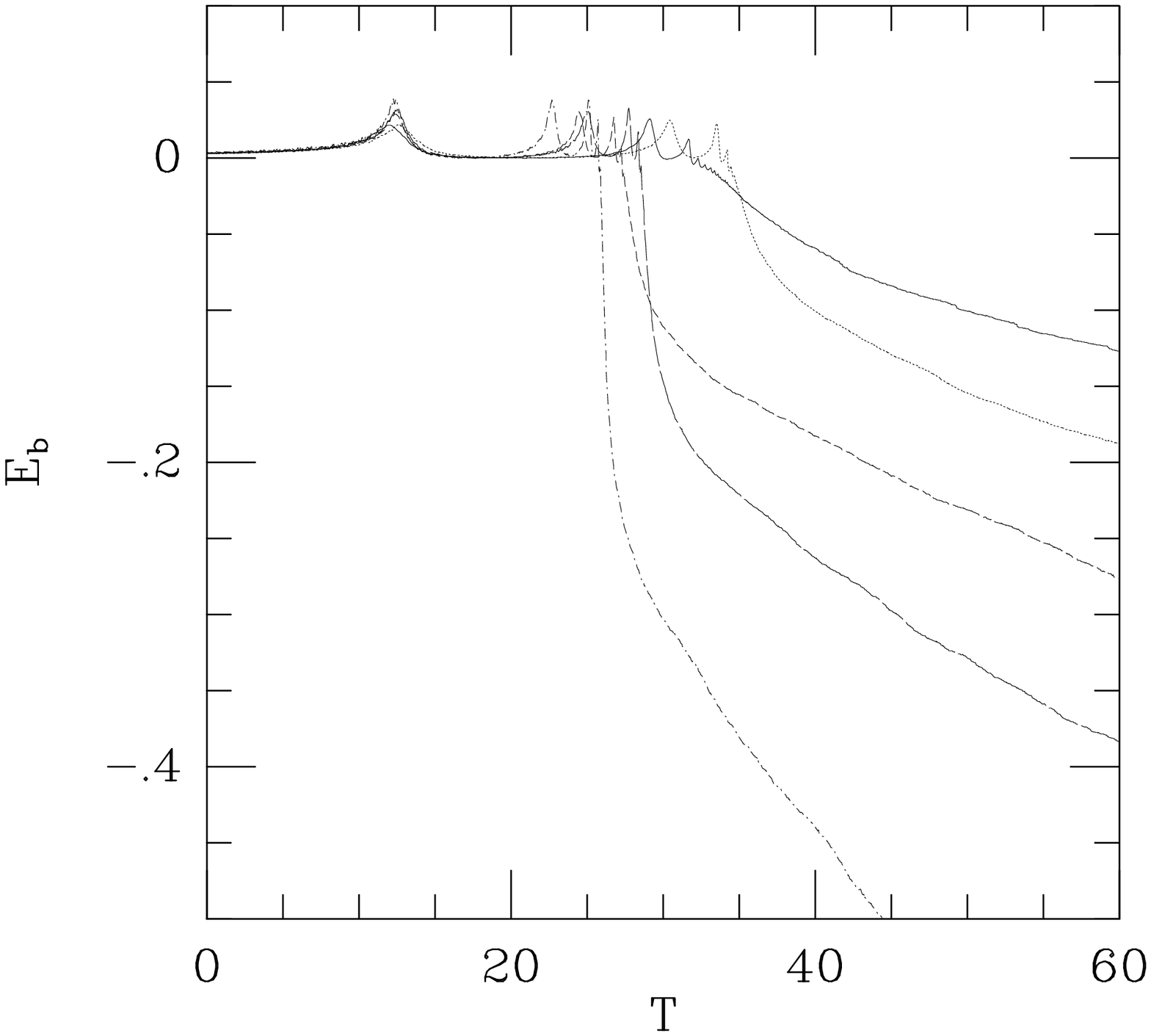]{
Same as figure 1 but for repeated mergers from N=256k run. Solid,
dotted, short-dashed, long-dashed and dot-dashed curves represent 1st
through 5th mergers, respectively.
\label{fig3}}

\figcaption[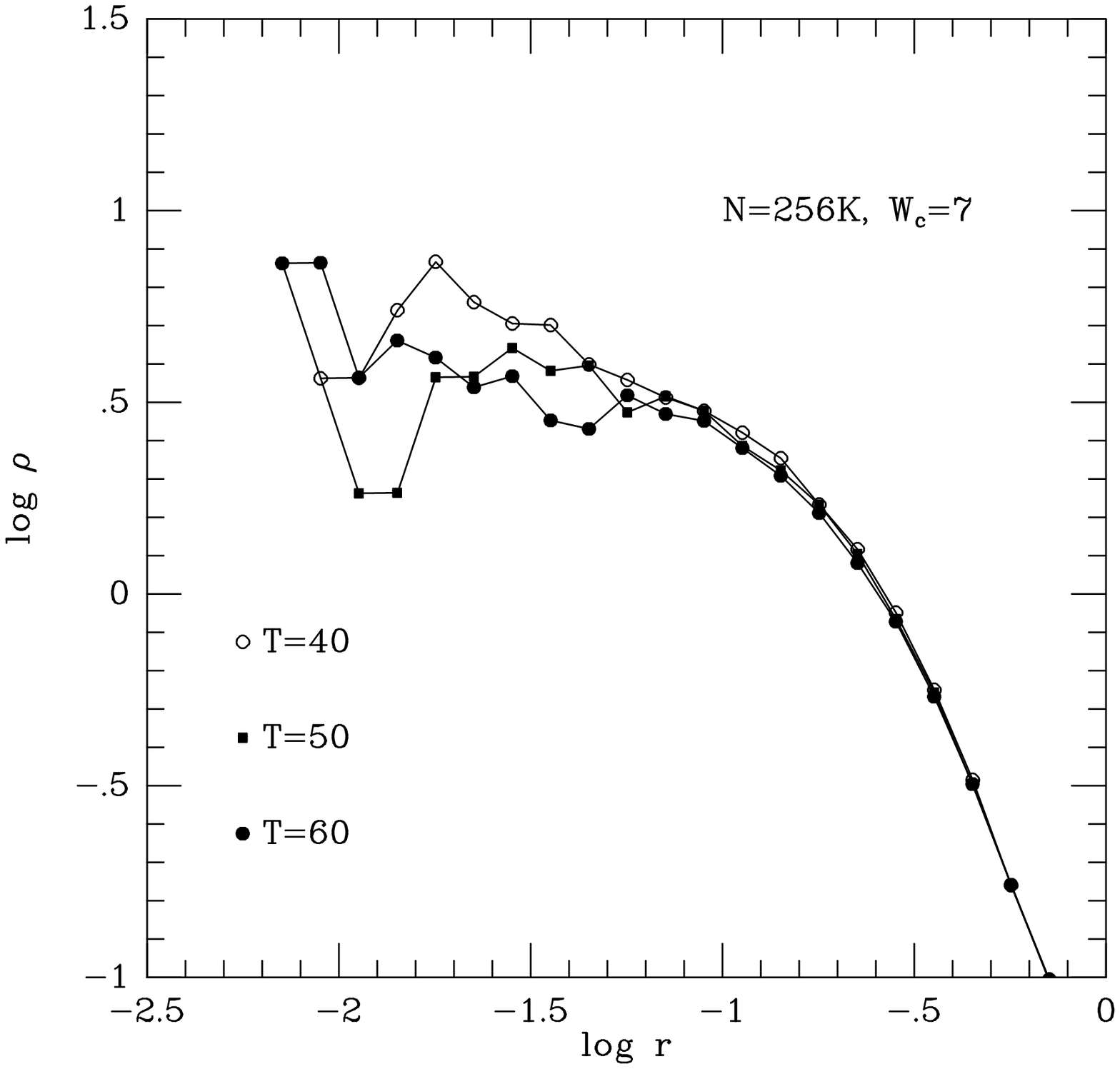,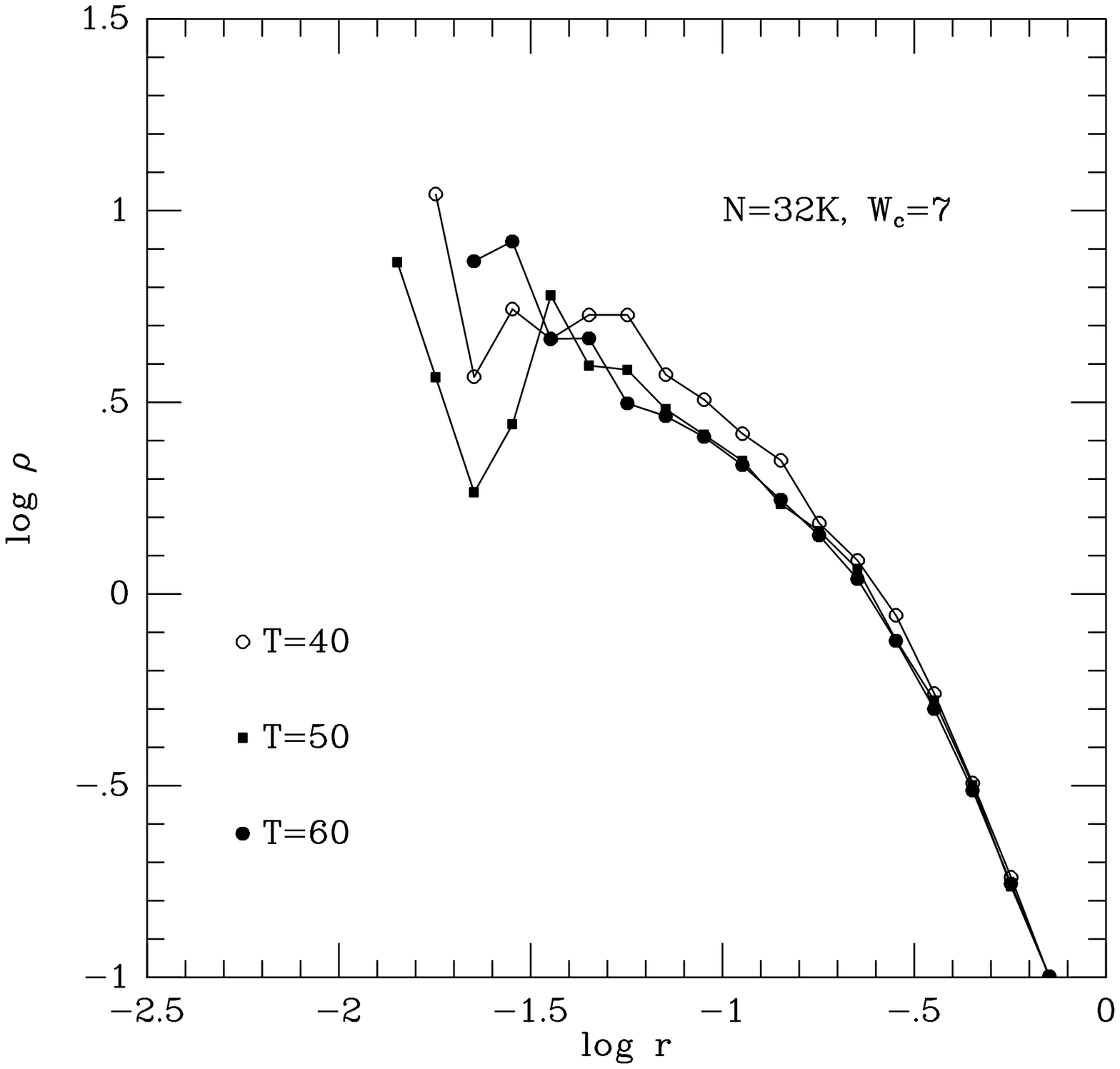,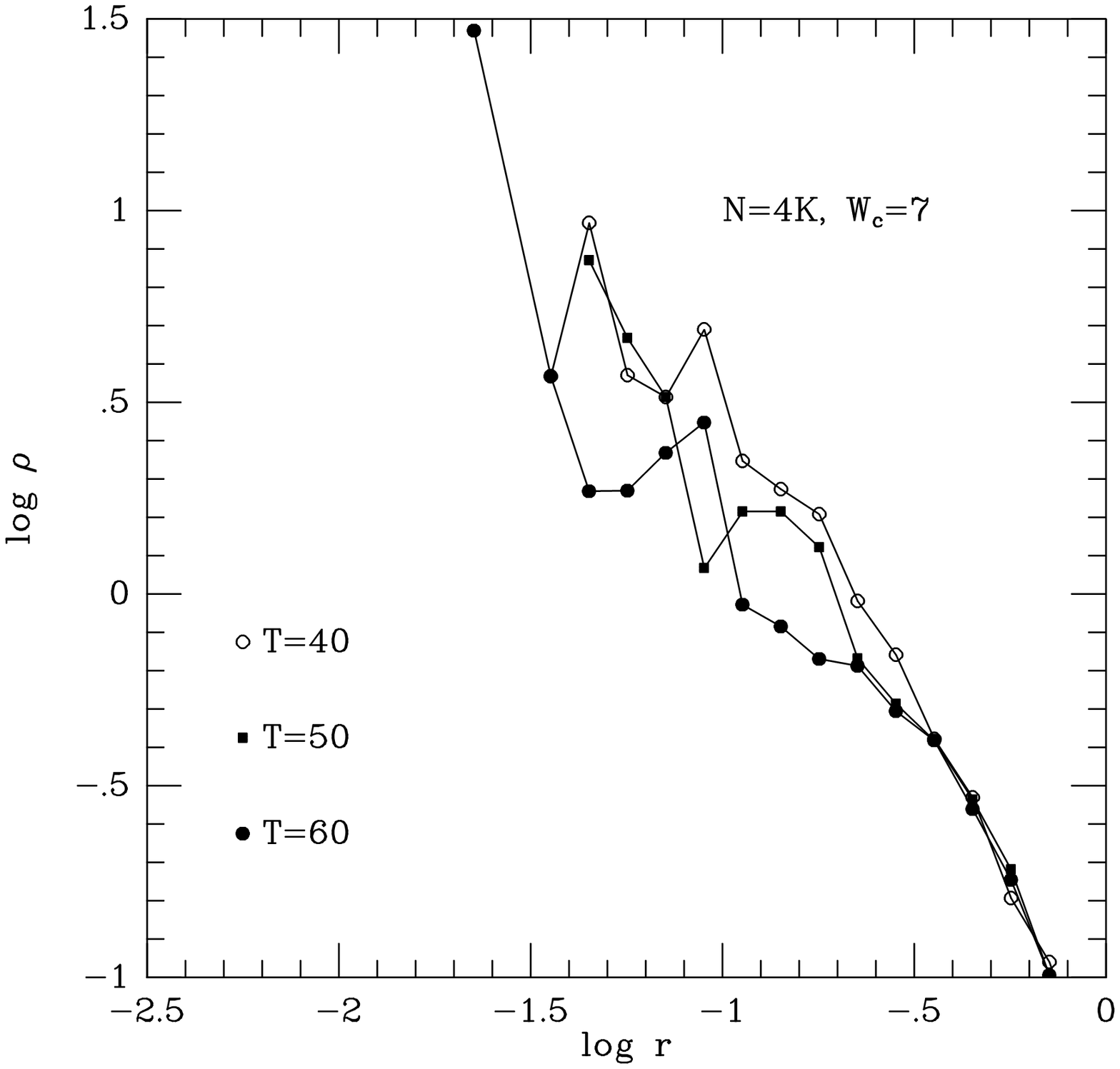,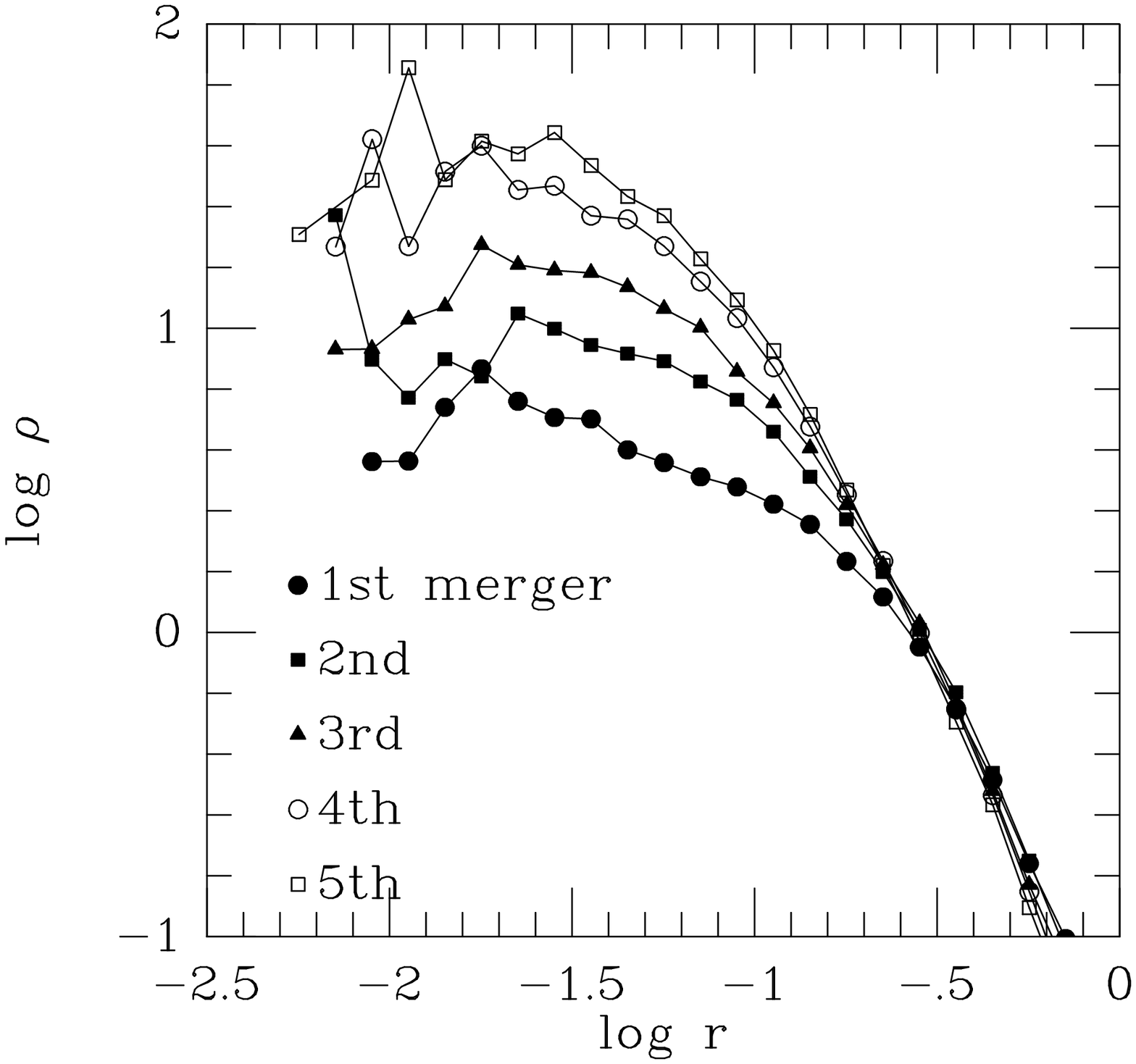]{
Density profiles of the central region of the mergers. (a-c) Runs
with (a) 256k, (b) 32k, and (c) 4k particles. Open circle, filled
squares and filled circles are the profiles at $t=40$, 50 and 60,
respectively. (d) Density profiles of repeated mergings.
Filled circles, filled squares,
filled triangles, open circles and open squares represent the profiles 
for 1st through 5th mergers. For all profiles, $t=40$. Initial total
number of particles is 256k. 
\label{fig4}}

\figcaption[bhn2afig5.eps]{
Same as figure 4 but for surface density profiles. 
\label{fig5}}

\clearpage

\plotone{bhn4fig1.eps}

\begin{center}
Figure 1
\end{center}

\plotone{bhn4fig2.eps}
\begin{center}
Figure 2
\end{center}

\plotone{bhn4fig3.eps}
\begin{center}
Figure 3
\end{center}

\plottwo{bhn4fig4a.eps}{bhn4fig4b.eps}

\plottwo{bhn4fig4c.eps}{bhn4fig4d.eps}

\begin{center}
Figure 4
\end{center}

\plottwo{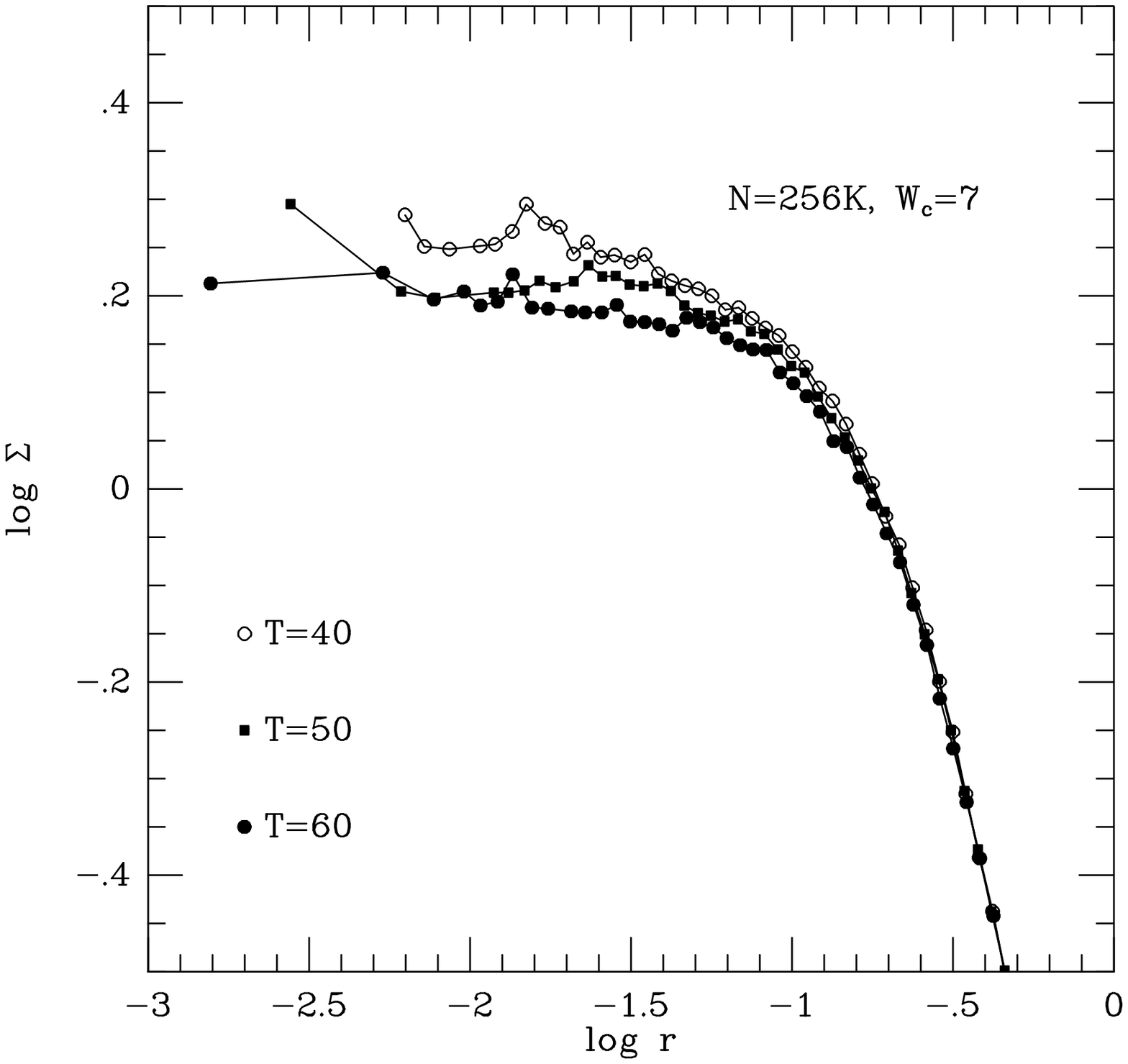}{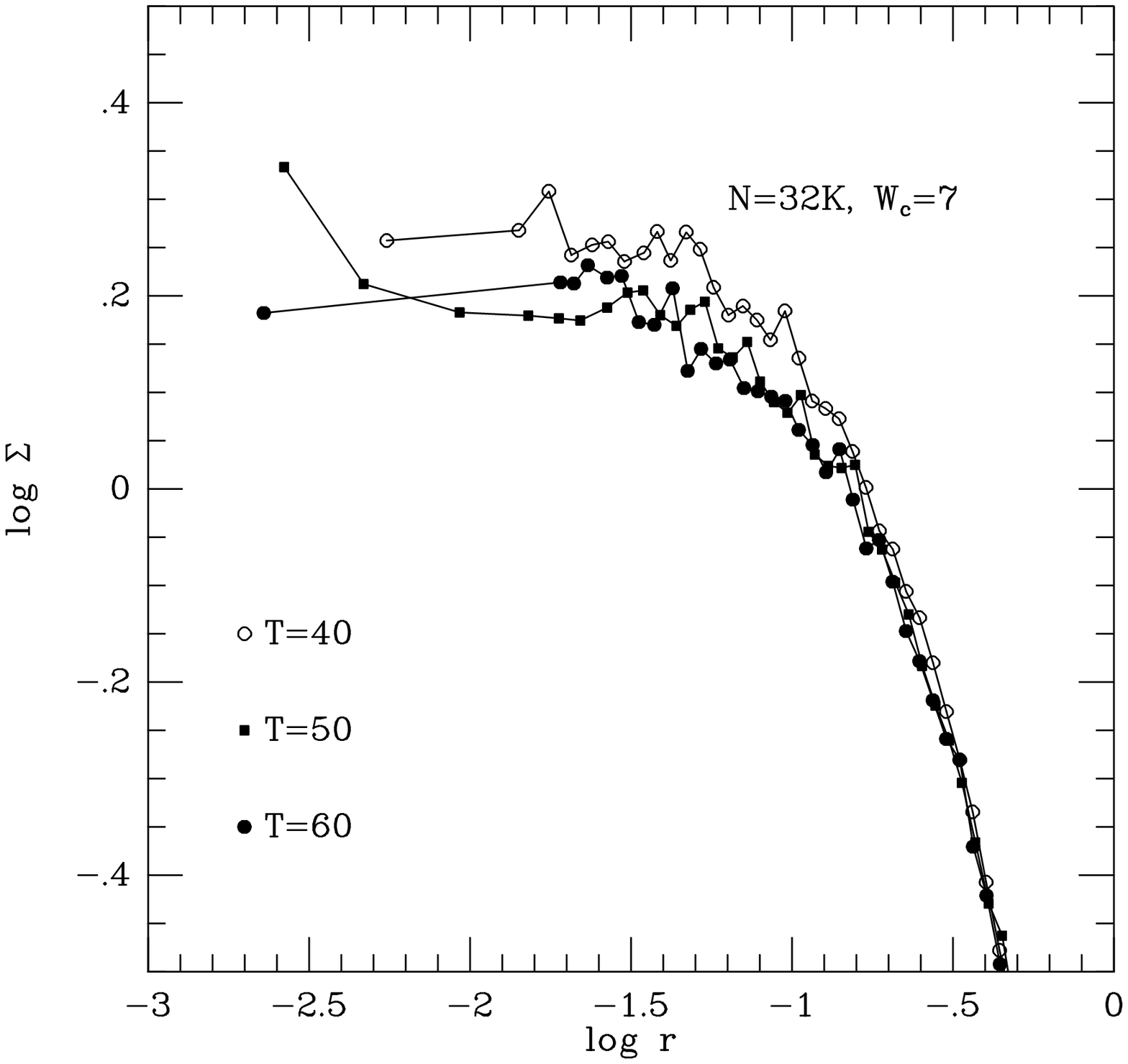}

\plotone{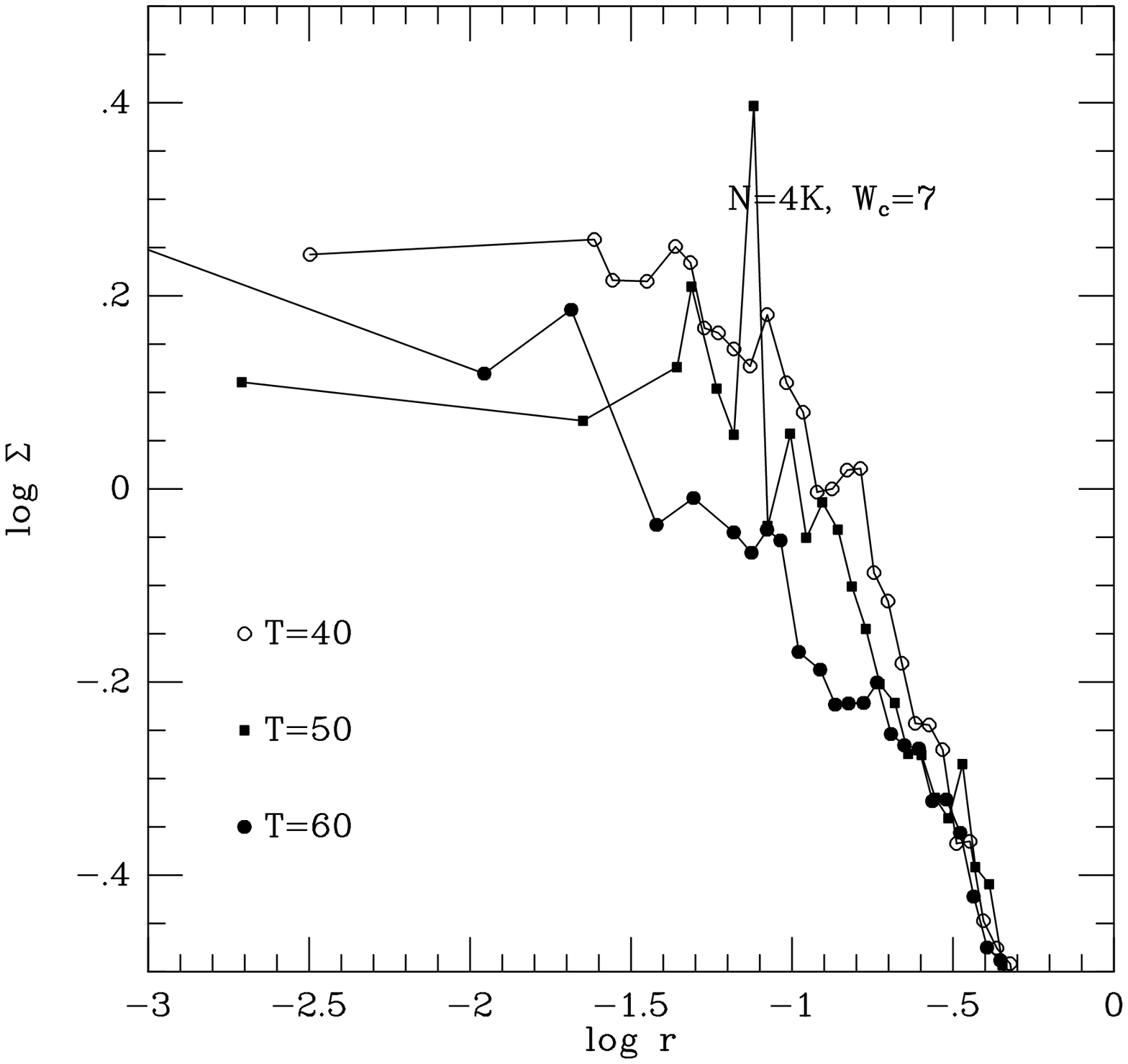}
\begin{center}
Figure 5
\end{center}

\end{document}